\documentclass[onecolumn]{emulateapj}
\usepackage{apjfonts}
\usepackage{amsbsy}
\usepackage{graphicx}
\epsfclipon

\newcommand{\ie}{{i.e.\hskip 3pt}}
\newcommand{\eg}{{e.g.,\hskip 3pt}}

\def\4he{$^4$He}
\def\3he{$^3$He}
\def\7li{$^7$Li}
\def\6li{$^6$Li}
\def\Yp{Y$_{\rm P}$~}
\def\yd{$y_{\rm D}$~}

\def\xie{$\xi_{e}$~}
\def\hii{H\thinspace{$\scriptstyle{\rm II}$}~}

\def\Nnu{N$_{\nu}$~}
\def\eten{$\eta_{10}$~}
\newcommand{\epm}{\ensuremath{e^{\pm}\;}}

\newcommand{\be}{\begin{equation}}
\newcommand{\ee}{\end{equation}}

\begin{document}


\title{Constraining The Universal Lepton Asymmetry}

\author{Vimal Simha\altaffilmark{1} and Gary Steigman\altaffilmark{2}}   

\altaffiltext{1}{Department of Astronomy, The Ohio State University, 140 
West 18th Ave., Columbus, OH 43210}
\altaffiltext{2}{Departments of Physics and Astronomy and Center for Cosmology 
and Astro-Particle Physics, The Ohio State University, 191 West Woodruff 
Ave., Columbus, OH 43210}


\begin{abstract}

The relic cosmic background neutrinos accompanying the cosmic microwave 
background (CMB) photons may hide a universal lepton asymmetry orders of 
magnitude larger than the universal baryon asymmetry.  At present, the 
only direct way to probe such an asymmetry is through its effect on the 
abundances of the light elements produced during primordial nucleosynthesis.  
The relic light element abundances also depend on the baryon asymmetry, 
parameterized by the baryon density parameter ($\eta_{\rm B} \equiv 
n_{\rm B}/n_{\gamma}$), and on the early-universe expansion rate, 
parameterized by the expansion rate factor ($S \equiv H'/H$) or, 
equivalently by the effective number of neutrinos (N$_{\nu} \equiv 3 + 
43(S^2 - 1)/7$). We use data from the CMB (and Large Scale 
Structure: LSS) along with the observationally-inferred relic abundances 
of deuterium and helium-4 to provide new bounds on the universal lepton 
asymmetry, finding
for $\eta_{\rm L}$, the analog of $\eta_{\rm B}$, 0.072$\pm$0.053 if
it is assumed that \Nnu = 3 and, 0.115$\pm$0.095 along with  
\Nnu = 3.3$^{+0.7}_{-0.6}$, if \Nnu is free to vary.

\end{abstract} 

\section{Introduction}

The standard models of particle physics and cosmology assume that in 
the early, radiation-dominated universe only the known, massless or 
light ($mc^2 \ll kT$) particles, including three flavors of light, 
active neutrinos (N$_{\nu} = 3$), contribute to energy density driving 
the universal expansion ($H^2 = 8\pi G\rho/3$; $\rho = \rho_{\rm R}$).  
It is also generally assumed that any universal lepton asymmetry is 
very small, of order the baryon asymmetry.\footnote{Charge conservation 
ensures that the very small proton-antiproton asymmetry, of order 
$\eta_{\rm B}$, is balanced by a correspondingly small asymmetry 
between electrons and positrons but, there are no such constraints 
on the size of any asymmetry between neutrinos and antineutrinos.}  
In analogy with the measure of the baryon asymmetry provided by 
$\eta_{\rm B}$, an asymmetry between neutrinos and antineutrinos of 
flavor $\alpha$ ($\alpha$ = e, $\mu$, $\tau$) can be described in terms 
of the neutrino chemical potential $\mu_{\alpha}$ or, in terms of the 
dimensionless degeneracy parameter $\xi_{\alpha} \equiv \mu_{\alpha}/kT$ 
by,
\be
\eta_{L} \equiv \Sigma_{\alpha} {n_{\nu_{\alpha}} - n_{\bar{\nu}_{\alpha}} 
\over n_{\gamma}} = {\pi^{3} \over 12\zeta(3)}\Sigma_{\alpha} \left[\left({\xi_{\alpha} 
\over \pi}\right) + \left({\xi_{\alpha} \over \pi}\right)^{3}\right].
\ee

For Big Bang Nucleosynthesis (BBN) the electron neutrinos play a key 
role through their charged-current weak interactions, which help to 
regulate the neutron-to-proton ratio ($p + e^{-} \leftrightarrow n + 
\nu_{e}$, $n + e^{+} \leftrightarrow p + \bar\nu_{e}$, $n \leftrightarrow
p + e^{-} + \bar\nu_{e}$).  Since the BBN-predicted \4he abundance is, 
to a very good approximation, determined by the neutron-to proton ratio 
at BBN, changes from the standard model value of this ratio will be 
reflected in its primordial abundance and, to a lesser extent, in the 
abundances of the other light elements produced during BBN.  These relic 
abundances therefore provide probes of a universal lepton asymmetry.  
For further discussion and previous analyses, see \eg \citealt{wfh, reeves, 
yb, bg, by,sch79, dr, scherrer, freese, terasawa, boes85, kang, kohri, esposito, 
ichi02, barger, kneller04, cuoco, serpico, steigman07, popa07, popa08}.

There is another way in which a significant neutrino degeneracy may
influence the early evolution of the Universe.  In the standard models 
of particle physics and cosmology, the energy density at, or just prior 
to, BBN is contributed by the cosmic background radiation photons, \epm 
pairs, and three flavors of extremely relativistic neutrinos.  In the
standard cosmology, the neutrinos constitute ~40\% of the total energy
density.  Any modification of the early-Universe energy density (or
expansion rate; see \eg \citealt{vs08}) can be parameterized in terms of the 
``effective" number of neutrinos by $\rho \rightarrow \rho' \equiv \rho 
+ \Delta$N$_{\nu}\rho_{\nu}$.  For the standard models, \Nnu = 3 at BBN, 
while in the post \epm annihilation epoch probed by the CMB, \Nnu = 3.046 
\citep{mangano05}.  A secondary effect of neutrino degeneracy is to increase 
the energy density in relativistic neutrinos predicted by the standard model, 
so that \Nnu $\rightarrow$ N$_{\nu} + \Sigma_{\alpha} \Delta$N$_{\nu}(\xi_{\alpha})$, 
where
\be
\Delta{\rm N}_{\nu}(\xi_{\alpha}) \equiv {30 \over 7}\left({\xi_{\alpha} \over \pi}\right)^{2}
+ {15 \over 7}\left({\xi_{\alpha} \over \pi}\right)^{4}.
\ee

In general, if the three active neutrino flavors mix only with each
other, neutrino oscillations ensure that their degeneracies equilibrate 
prior to BBN \citep{LS01,dolgov02,wong02,abb02}.  In the following we will 
assume this is the case and use \xie = $\xi_{\mu} = \xi_{\tau}$.  As a 
result,
\be
\eta_{\rm L} = {\pi^{3} \over 4\zeta(3)}\left[\left({\xi_{e} \over \pi}\right) + \left({\xi_{e}
\over \pi}\right)^{3}\right],
\ee
and
\be
\Sigma_{\alpha} \Delta{\rm N}_{\nu}\left(\xi_{\alpha}\right) = {90 \over 7}\left({\xi_{e} 
\over \pi}\right)^{2} + {45 \over 7}\left({\xi_{e} \over \pi}\right)^{4}.
\ee
Notice that, if $\xi_{e} = \xi_{\mu} = \xi_{\tau}$, then for $|\xi_{e}| 
\la 0.2$, $\Sigma_{\alpha} \Delta$N$_{\nu}(\xi_{\alpha}) \la 0.05$.  
Later, in \S6, we will relax this assumption so that $\xi_{e} \neq 
\xi_{\mu}$ and/or $\xi_{\tau}$)].  The only effect of non-zero values 
of $\xi_{\mu}$ and/or $\xi_{\tau}$ is to change the effective value 
of N$_{\nu}$, while non-zero values of \xie also contribute to \Nnu 
and, more importantly, they modify the neutron-to-proton ratio at 
BBN.

\section{Neutrino Degeneracy And BBN}

The stage is being set for BBN when the Universe is $\la$ 0.1 seconds 
old and the temperature is $\ga$ a few MeV, at which time the 
neutral-current weak interactions ($e^{+} + e^{-} \leftrightarrow 
\nu_{\alpha} + \bar{\nu}_{\alpha}$; $\alpha \equiv e, \mu, \tau$) are 
sufficiently rapid to maintain the neutrinos in equilibrium with the 
photon-\epm plasma.  When the temperature drops below $\sim 2$~MeV, the 
neutrinos begin to decouple from the photon-\epm plasma.  However, the
electron neutrinos continue to interact with the neutrons and protons 
through their charged-current weak interactions ($n + \nu_{e} 
\leftrightarrow p + e^{-},~p + \bar{\nu}_{e} \leftrightarrow n + 
e^{+}$,~$n \leftrightarrow p + e^{-} + \bar{\nu}_{e}$), enabling the 
neutron-to-proton ratio to track its equilibrium value of $n/p =~$exp$
(-\Delta m/kT)$, where $\Delta m$ is the neutron-proton mass difference.  
In the presence of a neutrino-antineutrino asymmetry for the electron
neutrinos, where $\xi_{e} \equiv \xi(\nu_{e}) = -\xi(\bar{\nu}_{e})$, 
the neutron-to-proton equilibrium ratio is shifted to $n/p =~$exp$(-\Delta 
m/kT-\xi_{e})$ which, for $\xi_{e} > 0$ results in fewer neutrons and 
less \4he.  

While the BBN-predicted abundances of D, \3he, and \7li are most 
sensitive to the baryon density, that of \4he is very sensitive to 
the neutron abundance when BBN begins and, therefore, to any 
e-neutrino degeneracy.  The primordial abundances of D, \3he, or \7li 
are baryometers, constraining $\eta_{\rm B}$, while the \4he primordial 
mass fraction (Y$_{\rm P}$) is a leptometer, constraining $\xi_{e}$.  
In the standard model of particle physics and cosmology with three 
species of neutrinos and their respective antineutrinos, the primordial 
element abundances depend on only one free parameter, the baryon 
density parameter, the post-\epm annihilation ratio (by number) of 
baryons to photons, $\eta_{\rm B}$ = $n_{\rm B}/n_{\gamma}$.  This 
parameter is related to baryon mass density parameter, $\Omega_{\rm B}$, 
the present-Universe ratio of the baryon mass density to the critical 
mass-energy density (see \citealt{steigman06}) by
\be
\eta_{10} = 10^{10}~n_{\rm B}/n_{\gamma} = 273.9~\Omega_{\rm B}h^2.
\ee
The abundance of \4he is sensitive to \xie and to the early universe 
expansion rate $S \equiv H'/H = (\rho'/\rho)^{1/2} = (1 + 
7\Delta$N$_{\nu}/43)^{1/2}$ \citep{steigman07}.  In contrast, the 
abundance of \4he is relatively insensitive to the baryon density 
since, to first order, all neutrons available when BBN begins are 
rapidly converted to \4he.  To a very good approximation 
\citep{kneller04,steigman07},
\be
{\rm Y}_{\rm P} = (0.2482 \pm 0.0006) + 0.0016(\eta_{\rm He} - 6),
\label{y_p}
\ee
where
\be
\eta_{\rm He} \equiv \eta_{10} + 100(S - 1) - 574\xi_{e}/4.
\ee
In eq.~\ref{y_p}, the effect of incomplete neutrino decoupling on the \4he 
mass fraction is accounted for according to the results of~\cite{mangano05}. 

In contrast to \4he, since the primordial abundances of D, \3he 
and \7li are set by the competition between two body production and 
destruction rates, they are more sensitive to the baryon density than 
to a non-zero lepton asymmetry or to a non-standard expansion rate.  
For a primordial ratio of D to H by number, \yd $\equiv 10^{5}({\rm 
D}/{\rm H})_{\rm P}$, in the range, $2 \la y_{\rm D} \la 4$, to a 
very good approximation \citep{kneller04, steigman07},
\be
y_{\rm D} = 2.64(1 \pm 0.03)(6/\eta_{\rm D})^{1.6},
\ee
where
\be
\eta_{\rm D} \equiv \eta_{10} - 6(S - 1) + 5\xi_{e}/4.
\ee
The effect of incomplete neutrino decoupling on this prediction is
at the $\sim 0.3$\% level \citep{mangano05}, about ten times smaller 
than the overall error estimate.

A good fit to the primordial ratio of \7li to H by number, $y_{\rm Li} 
\equiv 10^{10}({\rm Li}/{\rm H})_{\rm P}$ is provided by \citep{kneller04,
steigman07},
\be
y_{\rm Li} = 4.24(1 \pm 0.1)(\eta_{\rm Li}/6)^{2},
\ee
where
\be
\eta_{\rm Li} \equiv \eta_{10} - 3(S - 1) - 7\xi_{e}/4.
\ee

In our previous paper \citep{vs08} we assumed that \xie = 0 and concentrated 
on the constraints on $\eta_{10}$ and \Nnu which follow from BBN and the 
CMB (supplemented by large scale structure (LSS) data).  Here we will 
first set \Nnu = 3 ($S = 1$) and repeat our analysis for $\eta_{10}$ and 
$\xi_{e}$.  For consistency, in this case we will need to confirm that 
$|\xi_{e}|$ is bounded to be sufficiently small to justify the assumption 
that N$_{\nu} = 3 + \Sigma_{\alpha} \Delta$N$_{\nu}(\xi_{\alpha}) \approx 
3$.  Then, we will relax the assumption that \Nnu = 3 and use a combined 
BBN/CMB/LSS analysis to constrain all three parameters.  In this case, 
any contribution to $\Sigma_{\alpha} \Delta$N$_{\nu}(\xi_{\alpha})$ is 
automatically accounted for in our self-consistent determination of 
N$_{\nu}$.  

Before discussing our results, the primordial abundances adopted for 
our analysis are outlined.

\section{Adopted Primordial Abundances}

Since deuterium is destroyed as gas is cycled through stars, the deuterium
abundance inferred from high redshift (\ie young), low metallicity (\ie 
little stellar processing) QSO absorption line systems should provide 
the best estimate of its primordial abundance.  The weighted mean of 
the seven, high redshift, low metallicity D/H ratios from \citet{bt97},
\citet{bt98}, \citet{pb01}, \citet{omeara01}, \citet{kirkman03}, \citet{omeara06} 
and, most recently, from \citet{pettini08} is 
\be
y_{\rm D}=2.70^{+0.22}_{-0.20}.
\ee
For this relic abundance,
\be
\eta_{\rm D} = 5.92^{+0.30}_{-0.33}.
\ee

Since the post-BBN evolution of \4he is also monotonic, with its mass 
fraction, Y, increasing along with increasing metallicity, at low 
metallicity, the \4he abundance should approach its primordial value 
Y$_{\rm P}$.  Observations of helium and hydrogen recombination lines 
from low-metallicity, extragalactic \hii regions are of most value in 
determining Y$_{\rm P}$.  At present, corrections for systematic 
uncertainties dominate the estimates of the observationally-inferred 
\4he primordial mass fraction and, especially, its error.  Following 
\citet{steigman07,vs08}, we adopt for our estimate here,
\be
{\rm Y}_{\rm P}=0.240\pm0.006.
\ee
In this case,
\be
\eta_{\rm He} = 0.88\pm3.75.
\ee

While the central value of \Yp adopted here is low, the 
conservatively-estimated uncertainty is relatively large (some 
ten times larger than the uncertainty in the BBN-predicted 
abundance for a fixed baryon density).  In this context, it
should be noted that although very careful studies of the 
systematic errors in very limited samples of \hii regions 
provide poor estimators of \Yp as a result of their uncertain 
and/or model-dependent extrapolation to zero metallicity, 
they are of value in providing a robust {\it upper bound} to 
\Yp.  Using the results of \cite{os}, \cite{fk}, and \cite{peimbert07}, 
we follow \citet{steigman07,vs08} in adopting a $\sim 2\sigma$ 
upper bound of \Yp $\leq 0.255$.  This upper bound to \Yp corresponds
to an upper bound to $\eta_{\rm He} \leq 10.25$.

Although the BBN-predicted relic abundance of lithium provides a potentially 
sensitive baryometer ((Li/H)~$\propto \eta_{10}^{2}$, for $\eta_{10} 
\ga 4$), its post-BBN evolution is complex and model-dependent.  
In addition, lithium is only observed in the Galaxy, in its oldest, 
most metal-poor stars in galactic globular clusters and in the halo.  
However, these oldest galactic stars have had the most time to modify 
their surface lithium abundances, leading to the possibility that the 
observationally-inferred abundances may require large, uncertain, 
model-dependent corrections in order to use them to infer the primordial 
abundance of \7li.  In the absence of corrections for depletion, dilution, 
or gravitational settling, the data of \citet{ryan} and \citet{asplund06} 
suggest
\be
[{\rm Li}]_{\rm P} \equiv 12+{\rm log (Li/H)} = 2.1\pm0.1.
\ee
In contrast, in an attempt to correct for evolution of the surface 
lithium abundances, \citet{korn06} use their observations of a small, 
selected sample of stars in the globular cluster NGC6397, along with 
stellar evolution models which include the effect of gravitational 
settling to infer
\be
[{\rm Li}]_{\rm P} = 2.5\pm0.1.
\ee

In the following analysis the primordial abundances of D and \4he 
adopted here are used, along with CMB/LSS data, to estimate $\eta_{10}$, 
$\xi_{e}$, and $\Delta$N$_{\nu}$.  Then, using the inferred best values 
and uncertainties in these three parameters, the corresponding 
BBN-predicted abundance of \7li is derived and compared to its 
observationally inferred value.

\section{Neutrino Degeneracy ($\xi \neq 0$); Standard Expansion Rate (\Nnu = 3)}


\begin{figure}
\centerline{\epsfxsize=5truein\epsffile{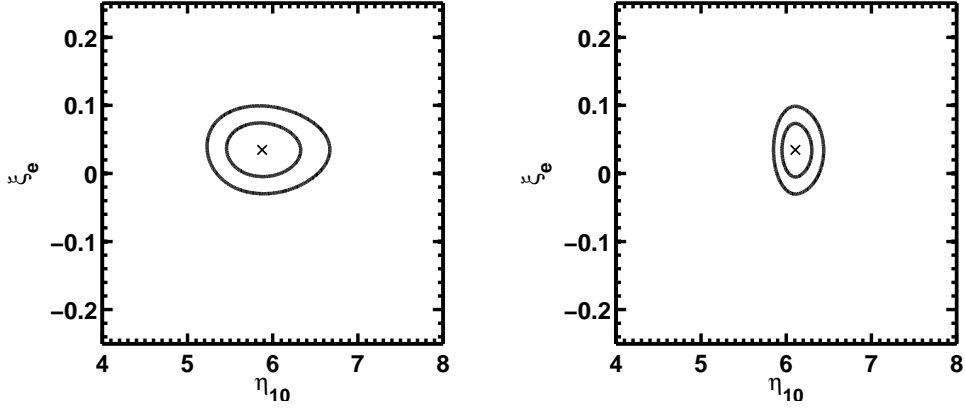}}
\caption{In the left hand panel the 68\% and 95\% contours in the $\xi -
\eta_{10}$ plane from BBN and the adopted relic abundances of D and 
\4he.  In the right hand panel, the BBN contraints convolved with 
the CMB/LSS constraint on $\eta_{10}$.  
}
\label{fig:etaxi}
\end{figure}


If attention is restricted to BBN alone, then there are two parameters 
inferred from observables, $\{\eta_{\rm D},\eta_{\rm He}\}$, and two 
unknowns, $\{\eta_{10},\xi_{e}\}$.  These are related by
\be
\xi_{e} = 4(\eta_{\rm D} - \eta_{\rm He})/579
\ee
and,
\be
\eta_{10} = (574\eta_{\rm D} + 5\eta_{\rm He})/579.
\ee
In the left hand panel of Figure \ref{fig:etaxi} are shown the 68\% and 
95\% contours in the \xie - $\eta_{10}$ plane derived from the adopted 
BBN abundances.  Notice that \xie and $\eta_{10}$ are virtually uncorrelated.  
From this approach it is found that $\xi_{e} = 0.035\pm0.026$ and, that 
$\eta_{10} = 5.88^{+0.30}_{-0.33}$.  The CMB/LSS provide an independent 
constraint on the baryon density, $\eta_{10} = 6.14^{+0.16}_{-0.11}$ 
\citep{vs08}.  Since these two estimates for $\eta_{10}$ are in agreement, 
we may use the CMB/LSS data to further restrict the allowed parameter 
space in the \xie - $\eta_{10}$ plane, as shown in the right hand panel 
of Figure \ref{fig:etaxi}. In this case, the value of \xie is unchanged 
and \eten$ = 6.11^{+0.12}_{-0.11}$.

Using these results for $\{\eta_{10},\xi_{e}\}$, the primordial abundance of 
\7li can be predicted using equations (10) and (11).  In this case we find 
that $\eta_{\rm Li} = 6.05^{+0.13}_{-0.12}$ and [Li] = 2.63$^{+0.04}_{-0.05}$.
The non-zero value of \xie which is consistent with BBN and the CMB is 
incapable of reconciling the BBN-predicted and observationally inferred 
primordial abundances of \7li.

An alternate approach which mixes BBN and the CMB provides another way 
to constrain the lepton asymmetry.  Since the \4he abundance is much 
more sensitive to \xie than is the deuterium abundance, equation (7) can 
be used along with the CMB value of $\eta_{10}$ to constrain $\xi_{e}$.  
This results in a virtually identical bound to that found from BBN alone, 
\xie = 0.037$\pm0.026$.  In contrast, combining the deuterium abundance with 
the CMB value of \eten using equation (9) gives \xie = -0.176$\pm0.272$, a 
range that is too broad to be of much value.

From these results we see that, at 95\% confidence, $|\xi_{e}| \leq 0.09$,
so that $\Sigma_{\alpha} \Delta$N$_{\nu}(\xi_{\alpha}) \leq 0.01$.  Such 
a small neutrino degeneracy has a negligible effect on the universal expansion 
rate during radiation domination, confirming the validity of the assumption 
that $S = 1$ (\Nnu = 3).  The corresponding lepton asymmetry $\eta_{\rm L} 
= 0.072\pm0.053$, while ``small", is orders of magnitude larger than 
the universal baryon asymmetry ($\eta_{\rm B} = 10^{-10}\eta_{10} = 
6\times10^{-10}$).

\section{Neutrino Degeneracy ($\xi \neq 0$); Non-Standard Expansion 
Rate (\Nnu $\neq 3$)}


\begin{figure}
\centerline{\epsfxsize=5truein\epsffile{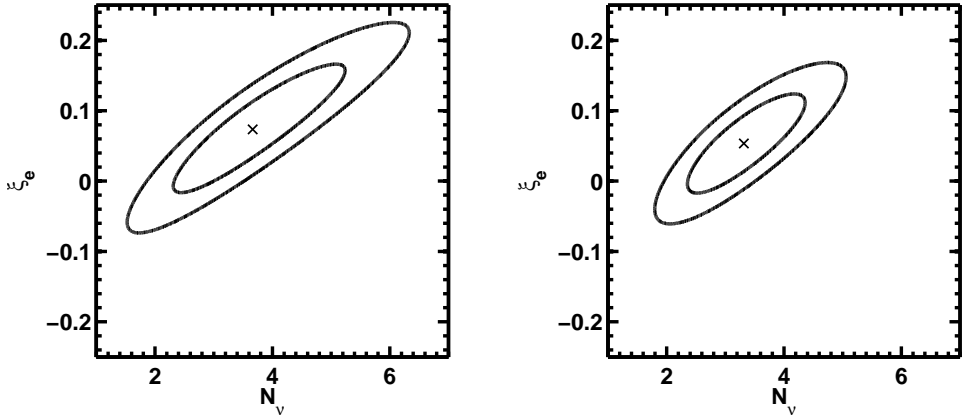}}
\caption{In the left hand panel the 68\% and 95\% contours in the $\xi_{e} 
-$ N$_{\nu}$ plane inferred from BBN and the adopted relic abundances 
of D and \4he, along with the CMB constraint on $\eta_{10}$.  In the 
right hand panel, the BBN contraints convolved with the CMB constraints 
on \Nnu and $\eta_{10}$.
}
\label{fig:nnuxi}
\end{figure}


In this case, with \Nnu and \xie free, there are three parameters 
to be determined: $\{\eta_{10},$N$_{\nu},\xi_{e}\}$, but only two 
{\it useful} relic abundances $\{y_{\rm D},$Y$_{\rm P}\}$.  To 
constrain these three parameters it is necessary to employ the 
CMB/LSS data along with that from BBN.  To obtain the best 
constraints a bit of care is required.  While the CMB/LSS provide 
a very good constraint on the baryon density parameter, they 
are less successful, at present, in constraining N$_{\nu}$.  
For example, \citet{vs08} find from the CMB/LSS data that while 
$\eta_{10} = 6.14^{+0.16}_{-0.11}$, \Nnu = 2.9$^{+1.0}_{-0.8}$ (or, 
$S-1 = -0.008^{+0.079}_{-0.068}$).  Such a large range in \Nnu will 
dilute the constraint on $\xi_{e}$, resulting in a less than optimal 
constraint on \xie given the available data.  However, as \citet{vs08} 
have noted, from the CMB/LSS the constraints on $\eta_{10}$ and \Nnu 
are virtually uncorrelated.  Therefore, in the following we adopt 
for our observational input $\{\eta_{10},\eta_{\rm D},\eta_{\rm He}\}$ 
and derive \Nnu and $\xi_{e}$ from them.  Since \xie is invisible to 
the CMB (except due to its contribution to the radiation energy 
density, which is accounted for in N$_{\nu}$), we may combine our
constraints on the $\{$N$_{\nu},\xi_{e}\}$ combination with the
independent constraint on \Nnu from the CMB, to further constrain 
this combination.

With $S$ (N$_{\nu}$) and \xie to be constrained from $\eta_{10}$,
$\eta_{\rm D}$, and $\eta_{\rm He}$, equations (7) and (9) may
be recast as
\be
S - 1 = (579\eta_{10} - 574\eta_{\rm D} - 5\eta_{\rm He})/2944
\ee
and,
\be
\xi_{e} = (106\eta_{10} - 100\eta_{\rm D} - 6\eta_{\rm He})/736.
\ee

In the left hand panel of Figure \ref{fig:nnuxi} are shown the 68\% and 95\% 
contours in the \xie - \Nnu plane derived from the adopted BBN abundances 
in combination with the constraint on \eten from the CMB/LSS.  From this 
approach we find $\xi_{e} = 0.073^{+0.056}_{-0.057}$ and \Nnu = 3.7$\pm0.9$.

The CMB/LSS provide an independent constraint on \Nnu, \Nnu = $2.9^{+1.0}_{-0.8}$ 
\citep{vs08}.  Since these two estimates are in agreement, the CMB/LSS data
may be used to further restrict the allowed parameter space in the \xie - 
\Nnu plane, as shown in the right hand panel of Figure \ref{fig:nnuxi}.  In 
this case, we obtain \xie = 0.056$\pm0.046$ and \Nnu = 3.3$^{+0.7}_{-0.6}$.

Using these results for $\{\eta_{10},{\rm N_{\nu}},\xi_{e}\}$, the primordial 
abundance of \7li can be predicted using equations (10) and (11).  In this 
case we find [Li] = $2.62^{+0.05}_{-0.06}$.  A non-zero value of \xie along 
with a non-standard expansion rate, consistent with BBN and the CMB, are 
still incapable of reconciling the BBN-predicted and observationally inferred 
primordial abundances of \7li. 

An alternate approach, mixing BBN and the CMB, provides another way to 
constrain the lepton asymmetry.  Since the \4he abundance is much more 
sensitive to \xie than is the deuterium abundance, equation (7) can be 
used along with the CMB/LSS values of $\eta_{10}$ and \Nnu to constrain 
$\xi_{e}$.  This results in a virtually identical bound, \xie = 0.053$\pm0.046$. 
In contrast, combining the deuterium abundance with the CMB value of 
\eten using equation (9) yields \xie $= -0.061^{+0.364}_{-0.376}$, whose 
range that is too broad to be of much value.

From these results, the effect of the neutrino degeneracy on the 
universal expansion rate during radiation domination may be computed, 
yielding $\Sigma_{\alpha} \Delta$N$_{\nu}(\xi_{\alpha}) \leq 0.03$ at 
95\% confidence. The corresponding lepton asymmetry, $\eta_{\rm L} = 
0.115\pm0.095$, while constrained to be ``small", may nonetheless be 
orders of magnitude larger than the universal baryon asymmetry 
($\eta_{\rm B} = 10^{-10}\eta_{10} = 6\times10^{-10}$).

\section{Discussion And Conclusions}


\begin{figure}
\centerline{\epsfxsize=5truein\epsffile{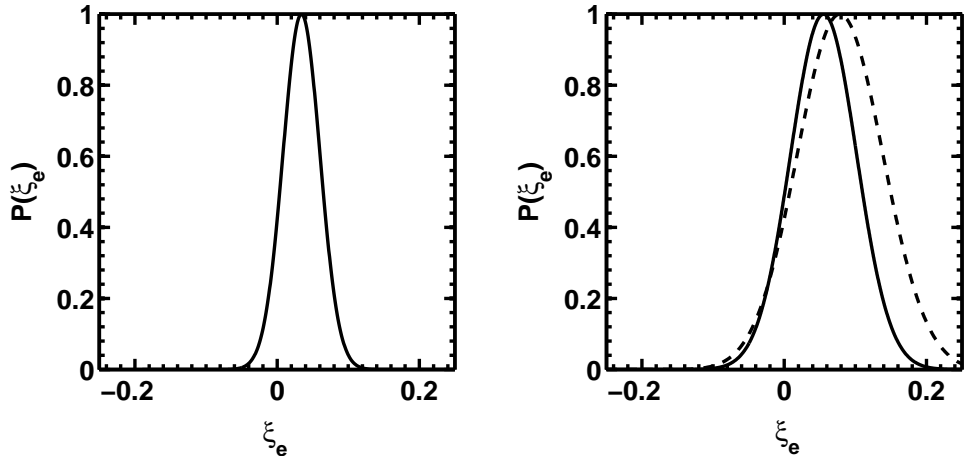}}
\caption{
The left hand panel shows the probability distribution for \xie, 
inferred from BBN and the adopted relic abundances of D and \4he, 
for the case where \Nnu = 3.  In the right hand panel, the dashed 
curve is the BBN constraint convolved with the CMB/LSS constraint 
on \eten alone, and the solid curve is the BBN constraint convolved 
with the CMB/LSS constraint on \eten and \Nnu.
}
\label{fig:xipdf}
\end{figure}

\begin{figure}
\centerline{\epsfxsize=2.5truein\epsffile{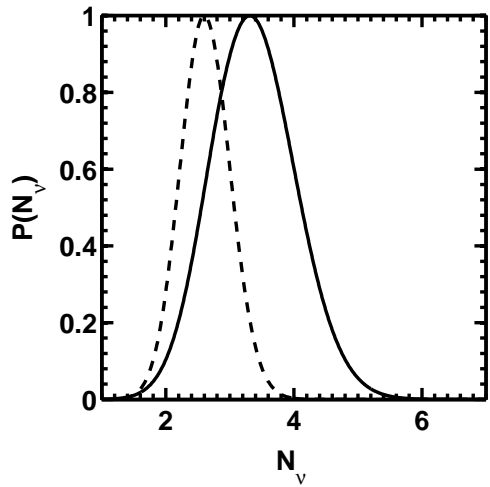}}
\caption{
The dashed curve shows the probability distribution for \Nnu for
\xie = 0 inferred from BBN and the relic abundances of D and \4he convolved
with the CMB/LSS constraints on \eten and \Nnu. The solid curve shows the same
for \xie $\neq$ 0.
}
\label{fig:nnupdf}
\end{figure}

The left hand panel of Figure \ref{fig:xipdf} shows the probability distribution 
of \xie for the standard expansion rate S = 0 (\Nnu = 3), derived after 
marginalizing over \eten, from BBN and the adopted primordial abundances of D 
and \4he.  The right hand panel of Figure \ref{fig:xipdf} shows the probability 
distribution of \xie for the more general case where a non-standard expansion 
rate S $\neq$ 0 (\Nnu $\neq$ 3) is allowed.  The constraints are based on 
combining BBN and the adopted primordial abundances of D and \4he with 
the independent constraints from the CMB/LSS.  Additional information from 
the CMB/LSS constraints on \eten are used before marginalizing over \eten and 
\Nnu to produce the dotted curve, while additional information from the CMB/LSS 
constraints on both \eten and \Nnu are used before marginalizing over \eten and 
\Nnu to produce the solid curve.

The dashed curve in Figure \ref{fig:nnupdf} shows the probability distribution of \Nnu
for \xie = 0, derived from BBN and the adopted primordial abundances of D and \4he with 
the independent constraints from the CMB/LSS after marginalizing over \eten. The solid curve
in Figure \ref{fig:nnupdf} shows the probability distribution of \Nnu for the more general 
case where a non-zero neutrino degeneracy \xie $\neq$ 0 is allowed. The constraints are based
on combining BBN and the adopted primordial abundances of D and \4he with 
the independent constraints from the CMB/LSS.  Additional information from 
the CMB/LSS constraints on \eten and \Nnu are used before marginalizing over \eten and 
\xie. Both of these constraints on \Nnu are consistent with each other and with the standard
model prediction of \Nnu $=$ 3.

Our results can be used to constrain any deviation in the universal expansion 
rate from its standard model value due to the increase in radiation energy
density from neutrino degeneracy.  Our constraint on the neutrino degeneracy 
leads to $\Sigma_{\alpha} \Delta$N$_{\nu}(\xi_{\alpha}) \leq 0.03$ at 95\% 
confidence.  In addition, the constraint on the neutrino degeneracy yields 
a corresponding constraint on any lepton asymmetry, $\eta_{\rm L}$: 
$\eta_{\rm L} = 0.072\pm0.053$ for \Nnu = 3 and $\eta_{\rm L} = 0.115\pm0.095$ 
when \Nnu $\neq$ 3. 

Using the constraints on $\eta_{10}$, N$_{\nu}$, and \xie, the BBN-predicted 
primordial abundance of \7li may be inferred.  For \Nnu = 3, [Li] = 
$2.63^{+0.04}_{-0.05}$ and for \Nnu $\neq 3$, [Li] = $2.62^{+0.05}_{-0.06}$.  
Both of these are considerably higher than the value ([Li]$_{\rm P} = 
2.1\pm 0.1$) determined from observations of metal-poor halo stars 
(\citet{ryan}, \citet{asplund06}) without any correction for depletion, 
destruction, or gravitational settling.  If, however, the correction 
proposed by \cite{korn06} is applied, the predicted and observed \7li 
abundances may, perhaps, be reconciled.  It remains an open question 
whether this lithium problem is best resolved by a better understanding 
of stellar physics or, if it is providing a hint of new physics beyond 
the standard model.    

In our analysis we have assumed that $\xi_{e} = \xi_{\mu} = \xi_{\tau}$.
Suppose instead that $\xi_{\mu} = \xi_{\tau} \neq \xi_{e}$ or, that $\xi_{e} 
= \xi_{\mu} \neq \xi_{\tau}$ \citep{dolgov04}. Since our constraints on \xie come from BBN, 
and they constrain \xie to be sufficiently small that the allowed degeneracy 
has minimal effect on the universal expansion rate ($\Delta$N$_{\nu}(\xi_{e}) 
\la 0.01$), the only way to probe non-zero values of $\xi_{\alpha} \equiv 
\xi_{\mu} \equiv \xi_{\tau}$ or $\xi_{\alpha} = \xi_{\tau}$, is through 
their effect on the expansion rate ($S$ or, $\Delta$N$_{\nu}$).  In these 
cases, it is possible that $\xi_{\alpha}$ may be $\gg \xi_{e}$.  For \Nnu 
= 3.3$^{+0.7}_{-0.6}$, $\Sigma_{\alpha} \Delta$N$_{\nu}(\xi_{\alpha}) \la 
1.7$ at $\sim 2\sigma$.  If it is assumed that $\xi_{\alpha} = \xi_{\mu} 
= \xi_{\tau}$, then $|\xi_{\alpha}| \la 2.34$ and $|\eta_{\rm L}| \la 5.0$.  
If, instead, it is assumed that $\xi_{\mu} = \xi_{e} \ll \xi_{\tau}$ (or, 
vice-versa for $\xi_{\mu}$), then $|\xi_{\tau}| \la 4.12$ and $|\eta_{\rm L}| 
\la 7.6$.

Of course, our results are sensitive to the relic abundances we have
adopted.  The simple but accurate fitting formulae \citep{kneller04} we 
have used make it easy to reevaluate our constraints for any adopted 
abundances.  The constraint on \xie is sensitive to the \4he abundance 
and is relatively insensitive to small changes in the D abundance.  For 
example, we repeated our analysis for a different primordial \4he mass 
fraction, ${\rm Y}_{\rm P} = 0.247\pm0.004$. This alternate abundance, 
in combination with the D abundance used in this paper, and the independent 
constraints on \eten and \Nnu from the CMB/LSS from \cite{vs08} yields \xie 
= 0.023$\pm$0.041.

Our results here are similar to, but considerably more restrictive than
those of \citet{barger} and of \citet{serpico}, due to the improved constraints 
on \Nnu and \eten from the WMAP 5-year and other CMB and LSS data.  The
analysis described here seems to be related to that in recent papers
by \citet{popa07,popa08}.  However, we fail to understand how they derive 
their constraints and why they find so much tighter bounds on \Nnu and
so much weaker bounds on $\xi_{e}$.

Except from its contribution to the radiation energy density and the
early Universe expansion rate, a lepton asymmetry in the neutrino sector 
is invisible to the CMB.  Future CMB experiments will improve the constraint 
on N$_{\nu}$ by measuring the neutrino anisotropic stress more accurately. 
According to \cite{bash04}, PLANCK should determine N$_{\nu}$ to an accuracy 
of $\sigma(N_{\nu}) \sim 0.24$ and CMBPOL, a satellite based polarization 
experiment, might improve it further to $\sigma(N_{\nu}) \sim 0.09$, 
independent of the BBN constraints.  Although still too large to provide a 
measure of the neutrino degeneracy, the tighter constraint on \Nnu can be 
used to further narrow the allowed ranges of \xie and \Nnu shown in Figure 
\ref{fig:nnuxi}.  

\begin{center}
ACKNOWLEDGMENTS
\end{center}
This research is supported at The Ohio State University by a grant 
(DE-FG02-91ER40690) from the US Department of Energy.  We thank J.
Beacom and G. Gelmini for useful discussions. 


\singlespace

\end{document}